\documentclass[11pt]{article}

\usepackage{polyglossia}
\setmainlanguage{english}

\usepackage[margin=1in]{geometry}
\usepackage{microtype}
\usepackage{graphicx,booktabs}
\usepackage{subcaption}
\usepackage{multirow}
\usepackage{float}
\usepackage{amsmath,amssymb}
\usepackage[hidelinks]{hyperref}
\usepackage{orcidlink}

\setotherlanguage{arabic}

\newfontfamily\arabicfont[
  Script=Arabic,
  Path=fonts/,
  UprightFont = *-Regular,
  BoldFont    = *-Bold
]{NotoNaskhArabic}

\newfontfamily\arabicfontsf[
  Script=Arabic,
  Path=fonts/,
  UprightFont = *-Regular,
  BoldFont    = *-Bold
]{NotoNaskhArabic}

\newcommand{\arbtxt}[1]{\textarabic{#1}}
\newcommand{\arb}{\arbtxt}

\DeclareMathOperator*{\softmax}{softmax}

\title{Arabic TTS with FastPitch: Reproducible Baselines, Adversarial Training, and Oversmoothing Analysis}
\author{Lars Nippert\,\orcidlink{0009-0009-3527-5032}}
\date{} 

\begin{document}
\maketitle
\begin{abstract}
\noindent
Arabic text-to-speech (TTS) remains challenging due to limited resources and complex phonological patterns. 
We present reproducible baselines for Arabic TTS built on the FastPitch architecture and introduce cepstral-domain metrics for analyzing oversmoothing in mel-spectrogram prediction. 
While traditional $L_p$ reconstruction losses yield smooth but over-averaged outputs, the proposed metrics reveal their temporal and spectral effects throughout training. 
To address this, we incorporate a lightweight adversarial spectrogram loss, which trains stably and substantially reduces oversmoothing. 
We further explore multi-speaker Arabic TTS by augmenting FastPitch with synthetic voices generated using XTTSv2, resulting in improved prosodic diversity without loss of stability.
The code, pretrained models, and training recipes are publicly available at:
\url{https://github.com/nipponjo/tts-arabic-pytorch}.
\end{abstract}

\section{Introduction}

End-to-end neural text-to-speech (TTS) systems have achieved remarkable 
naturalness and robustness in recent years. Architectures such as 
Tacotron~2~\cite{Shen2017NaturalTS} and FastPitch~\cite{Lancucki2020FastpitchPT} 
have become standard baselines for English and other high-resource languages, 
enabling expressive and controllable speech synthesis. Despite these advances, 
Arabic remains comparatively underexplored. The language poses unique 
challenges for TTS due to its complex morphology, optional diacritics, 
and wide dialectal variation. Furthermore, publicly available Arabic 
datasets are relatively scarce, limiting reproducibility and slowing 
progress compared to English and Chinese.

Most neural TTS models are trained with simple $L_p$ reconstruction losses 
applied to mel-spectrograms. While these losses are effective for convergence, 
they encourage average predictions, suppressing fine spectral detail and 
leading to the well-known problem of \emph{oversmoothing}. This results in 
muffled spectrograms and degraded perceptual quality. Adversarial objectives 
can alleviate this effect, but their use in Arabic TTS remains limited and 
their behavior has not been systematically studied. At the same time, 
there is a lack of established objective metrics that can quantify 
oversmoothing during training and allow principled model comparison.

In this work, we make the following contributions:
\begin{itemize}
  \item We provide reproducible baselines for Arabic TTS using FastPitch, trained on publicly available \emph{Arabic Speech Corpus} (ASC)~\cite{nawar2016phdthesis}.
  \item We introduce a set of cepstral-domain metrics for quantifying 
        oversmoothing in mel-spectrograms, which can be tracked 
        during training.
  \item We incorporate an adversarial spectrogram discriminator, inspired 
        by \emph{DeepFillv2}~\cite{yu2019freeformimageinpaintinggated}, and show that it runs stably 
        in our setup and significantly reduces oversmoothing.
  \item We extend FastPitch to a multi-speaker setting by augmenting 
        the training corpus with synthetic samples generated by \emph{XTTSv2}~\cite{casanova2024xtts}, 
        yielding models with three additional voices.
  \item We release all code, pretrained models, and training recipes to support reproducibility and further research.\footnote{Available at \url{https://github.com/nipponjo/tts-arabic-pytorch}.}
\end{itemize}

The remainder of this paper is structured as follows. 
Section~\ref{sec:methods} describes the datasets, preprocessing pipeline, model architectures, and training procedure. 
Section~\ref{sec:eval-smooth} introduces the proposed oversmoothing metrics. 
Section~\ref{sec:results} presents the experimental results, including adversarial training and multi-speaker extensions. The findings are discussed in 
Section~\ref{sec:discussion}, and Section~\ref{sec:conclusion} concludes the paper.

\section{Methods} \label{sec:methods}

This section describes the full experimental pipeline used to develop and evaluate our Arabic FastPitch TTS systems. 
We first summarize the datasets and preprocessing steps used to obtain phoneme sequences and mel-spectrogram targets (Sections~\ref{sec:data}–\ref{sec:text}). 
We then present the used models and training objectives, including the standard regression losses and the additional adversarial spectrogram discriminator (Sections~\ref{sec:models}–\ref{sec:losses}). 
To quantify oversmoothing, we introduce a set of cepstral-domain evaluation metrics designed to capture changes in fine spectral detail during training (Section~\ref{sec:eval-smooth}). 
Finally, we outline the evaluation protocol and training setup used across all experiments (Sections~\ref{sec:eval}–\ref{sec:training}). 
Together, these components provide a reproducible and fully specified framework for the analyses reported in Section~\ref{sec:results}.

\subsection{Data} \label{sec:data}

We use both natural and synthetic Arabic speech data: the Arabic Speech Corpus (ASC) for baseline training and XTTSv2-generated speech for multi-speaker experiments.

\subsubsection{Arabic Speech Corpus (ASC)} \label{sec:data-asc}
We use Nawar Halabi's \emph{Arabic Speech Corpus} (ASC)~\cite{nawar2016phdthesis}\footnote{Available at \url{https://en.arabicspeechcorpus.com/}.}, 
a publicly available Modern Standard Arabic dataset containing 
1,813 utterances (3~hours and 32~minutes of audio) spoken by a single male speaker 
with a south Levantine, Damascene accent.

Roughly half of the utterances are full sentences drawn from news-style 
text, such as 
\newline
\arb{أتَاحَتْ لِلبَائِعِ المُتَجَوِّلِ أنْ يَكُونَ جَاذِباً لِلمُوَاطِنِ الأَقَلِّ دَخْلاً} (\emph{'atāHat lilbā'i3i l-mutajawwili 'an yakūna jāḏiban lilmuwāṭini l-'aqalli daxlan}, “It allowed the street vendor to be attractive to the lower-income citizen.”), 
which provide natural prosody and contextual variation. 
The remaining utterances are are “phoneme-rich” constructions such as
\arb{تَأَّاصَوَّرَ وَتَأَّاصَرَ وَتُؤّاصَ تَصَأَّا} (\emph{ta''āSawwara wata''āSara watu''āSa taSa''ā}), 
which are not semantically meaningful but were specifically included to 
cover rare phonemes, stress patterns, and phonotactic contexts that are 
unlikely to appear frequently in spontaneous text. 

\begin{figure}[H]
  \centering
  \includegraphics[width=0.85\linewidth]{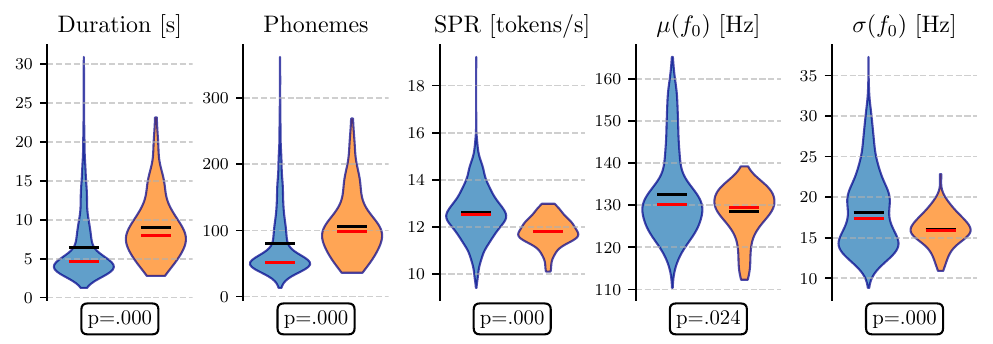}
  \caption{
  Distribution of utterance-level features in the ASC training set (blue) 
  and test set (orange). Each violin plot shows the full distribution, with 
  the red line indicating the median and the black line indicating the mean. 
  The training set contains many phoneme-rich pseudo-words designed to cover 
  the full phonetic inventory, whereas the test set consists exclusively of 
  natural sentences. This mismatch leads to systematic differences in duration, 
  number of phonemes, speaking rate, and pitch statistics, which should be 
  taken into account when interpreting evaluation metrics. 
  $p$-values from the \emph{Mann--Whitney U test}~\cite{mann1947test}.}
  \label{fig:train_test_violins}
\end{figure}

The Arabic Speech Corpus includes an official \emph{test set} consisting of 100 utterance (18~minutes of audio), which is used for model evaluation in our experiments. Unlike the training portion, which contains a large fraction of phoneme-rich pseudo-words designed to cover the phonetic inventory, the test set consists only of natural sentences. 
Consequently, several utterance-level statistics differ substantially between 
train and test (see Figure~\ref{fig:train_test_violins} and Table~\ref{tab:train_test_metrics}). These differences should be kept in mind when interpreting absolute values of the reported metrics, as they partly reflect corpus design rather than model behavior.

\begin{table}[h]
\centering
\caption{Comparison of utterance-level statistics for the ASC training set 
(1,813 utterances) and official test set (100 utterances). 
Values are reported as mean~$\pm$~standard deviation. 
$p$-values are computed using the \emph{Mann--Whitney U test}~\cite{mann1947test}.}
\label{tab:train_test_metrics}
\begin{tabular}{lllc}
\toprule
Measure & Train & Test & $p$-value \\
\midrule
Utterance duration [s] & $6.44 \pm 4.34$ & $8.99 \pm 4.05$ & $0.000$ \\
Phonemes per utterance & $80.3 \pm 53.5$ & $106.0 \pm 47.8$ & $0.000$ \\
Speaking rate (SPR) [tokens/s] & $12.6 \pm 1.1$ & $11.8 \pm 0.604$ & $0.000$\\
Mean pitch $\mu(f_0)$ [Hz] & $132.0 \pm 10.3$ & $128.0 \pm 5.95$ & $0.024$\\
Pitch standard deviation $\sigma(f_0)$ [Hz]  & $18.1 \pm 4.75$ & $16.0 \pm 2.05$ & $0.000$\\
\bottomrule
\end{tabular}
\end{table}

\subsubsection{Synthetic Data (XTTSv2)} \label{sec:data-synth}

To extend our experiments beyond the single-speaker setting, we generated three additional synthetic voices using the XTTSv2~\cite{casanova2024xtts} model. This allowed us to explore multi-speaker training while maintaining control over data comparability. Although in principle an arbitrary number of synthetic utterances could be sampled, we chose to resynthesize the official ASC training and test sets for each synthetic speaker. This ensures that all speakers share identical linguistic content, making differences in the metrics more directly attributable to speaker characteristics rather than to text distribution.

The resulting set consists of the original ASC voice (Speaker~0) and three additional voices (Speakers~1-3) synthesized from distinct speaker embeddings. Speaker~1 is based on male embeddings, while Speakers~2 and~3 are derived from female embeddings, allowing us to contrast male and female characteristics in both prosodic and cepstral metrics. Table~\ref{tab:speaker_metrics} summarizes these basic statistics across train and test splits.

Speaker 1 exhibits a lower mean pitch ($\approx$103 Hz) and reduced pitch variability, producing a flatter prosody compared to the ASC reference. 
By contrast, Speakers 2 and 3, generated from female embeddings, show substantially higher mean pitch values (225 Hz and 204 Hz, respectively) and greater variability. 

The cepstral-domain metrics reflect this difference: female voices exhibit higher HQER and CCentroid values. This is expected because the higher fundamental frequency in female speech shifts harmonic energy toward higher quefrencies, which systematically raises these measures—similar to how the \emph{cepstral peak prominence} (CPP)~\cite{hillenbrand1994cpp} is known to be higher in female voices.

\begin{table}[H]
\centering
\renewcommand{\arraystretch}{1.2}

\begin{subtable}[t]{0.48\textwidth}
\centering
\begin{tabular}{lcc}
\toprule
Metric & \multicolumn{2}{c}{Speaker 0} \\
\cmidrule(lr){2-3}
 & Train & Test \\
\midrule
Duration [s] & $6.44 \pm 4.34$ & $8.99 \pm 4.05$ \\
SPR [tokens/s] & $12.6 \pm 1.1$ & $11.8 \pm 0.604$ \\
$\mu(f_0)$ [Hz] & $132.0 \pm 10.3$ & $128.0 \pm 5.95$ \\
$\sigma(f_0)$ [Hz] & $18.1 \pm 4.75$ & $16.0 \pm 2.05$ \\
HQER [\%] & $14.0 \pm 2.43$ & $13.9 \pm 1.39$ \\
CSlope [dB/bin] & $-0.421 \pm 0.0254$ & $-0.406 \pm 0.0193$ \\
CCentroid [bin] & $4.72 \pm 0.6$ & $5.09 \pm 0.32$ \\
CRoll95 [bin] & $22.5 \pm 1.86$ & $23.8 \pm 1.04$ \\
\bottomrule
\end{tabular}
\caption{Speaker 0}
\end{subtable}%
\hfill
\begin{subtable}[t]{0.48\textwidth}
\centering
\begin{tabular}{cc}
\toprule
\multicolumn{2}{c}{Speaker 1} \\
\cmidrule(lr){1-2}
 Train & Test \\
\midrule
$4.84 \pm 3.08$ & $6.35 \pm 2.84$ \\
$16.4 \pm 1.48$ & $16.8 \pm 1.12$ \\
$105.0 \pm 3.61$ & $103.0 \pm 2.71$ \\
$11.0 \pm 2.67$ & $10.5 \pm 1.8$ \\
$11.8 \pm 2.29$ & $13.1 \pm 1.3$ \\
$-0.48 \pm 0.017$ & $-0.474 \pm 0.0134$ \\
$4.5 \pm 0.69$ & $4.96 \pm 0.368$ \\
$19.2 \pm 2.46$ & $20.6 \pm 1.42$ \\
\bottomrule
\end{tabular}
\caption{Speaker 1}
\end{subtable}

\vspace{1em}

\begin{subtable}[t]{0.48\textwidth}
\centering
\begin{tabular}{lcc}
\toprule
Metric & \multicolumn{2}{c}{Speaker 2} \\
\cmidrule(lr){2-3}
 & Train & Test \\
\midrule
Duration [s] & $5.81 \pm 3.27$ & $7.43 \pm 2.89$ \\
SPR [tokens/s] & $13.3 \pm 1.54$ & $14.0 \pm 1.36$ \\
$\mu(f_0)$ [Hz] & $228.0 \pm 8.71$ & $225.0 \pm 7.23$ \\
$\sigma(f_0)$ [Hz] & $34.9 \pm 4.16$ & $34.8 \pm 3.44$ \\
HQER [\%] & $26.8 \pm 3.94$ & $26.2 \pm 2.7$ \\
CSlope [dB/bin] & $-0.447 \pm 0.0197$ & $-0.443 \pm 0.0139$ \\
CCentroid [bin] & $6.52 \pm 0.732$ & $6.5 \pm 0.445$ \\
CRoll95 [bin] & $23.0 \pm 1.3$ & $23.3 \pm 0.834$ \\
\bottomrule
\end{tabular}
\caption{Speaker 2}
\end{subtable}%
\hfill
\begin{subtable}[t]{0.48\textwidth}
\centering
\begin{tabular}{cc}
\toprule
\multicolumn{2}{c}{Speaker 3} \\
\cmidrule(lr){1-2}
Train & Test \\
\midrule
$5.52 \pm 3.22$ & $6.94 \pm 2.89$ \\
$14.1 \pm 1.41$ & $15.1 \pm 1.15$ \\
$203.0 \pm 10.1$ & $204.0 \pm 6.85$ \\
$39.9 \pm 6.38$ & $39.1 \pm 5.28$ \\
$22.7 \pm 3.36$ & $23.9 \pm 2.49$ \\
$-0.432 \pm 0.0219$ & $-0.433 \pm 0.0156$ \\
$6.31 \pm 0.73$ & $6.66 \pm 0.465$ \\
$23.6 \pm 1.33$ & $24.1 \pm 0.91$ \\
\bottomrule
\end{tabular}
\caption{Speaker 3}
\end{subtable}

\caption{Comparison of prosodic and cepstral metrics (mean $\pm$ std) across four speakers. Speaker 0 is the original ASC voice, Speaker 1 is a synthetic male voice, while Speakers 2 and 3 are synthetic female voices. Values are reported separately for train and test subsets of the corpus.}
\label{tab:speaker_metrics}
\end{table}

\subsection{TTS Pipeline Overview} \label{sec:pipeline}

\begin{figure}[H]
  \centering
  \includegraphics[width=0.80\linewidth]{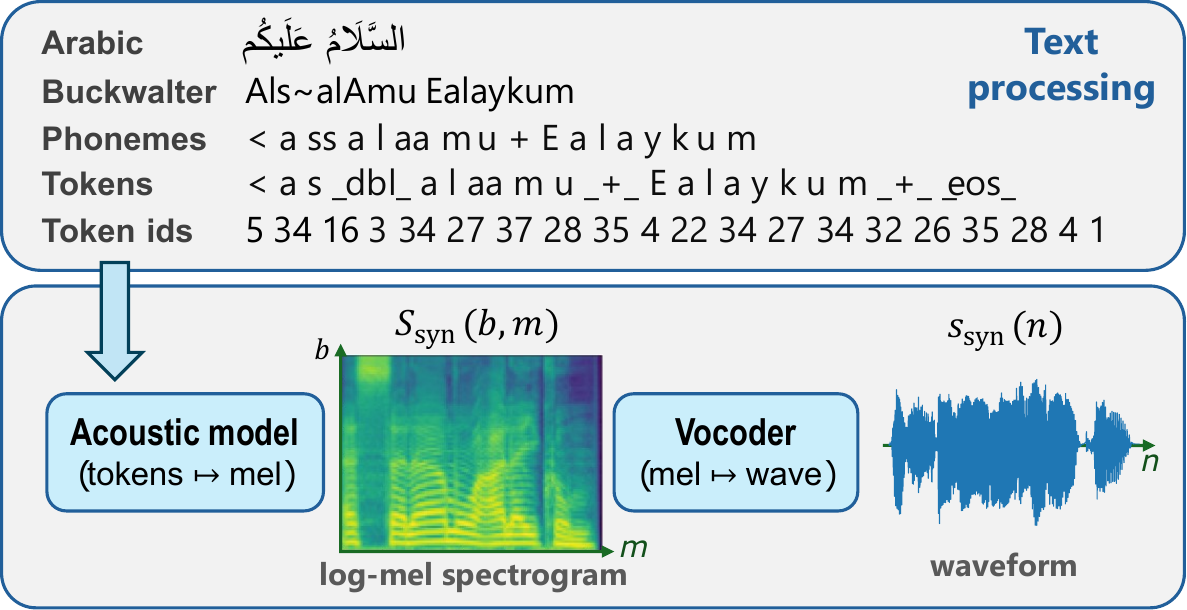}
  \caption{
  Overview of the Arabic TTS pipeline. 
  The input text is first transliterated and phonemized into a sequence of phonemes, which are mapped to token IDs. 
  The acoustic model predicts a log-mel spectrogram $S_{\text{syn}}(b,m)$ from the token sequence, and a neural vocoder converts it into the final waveform $s_{\text{syn}}(n)$. 
  The example illustrates the full processing chain from Arabic script to audio.
  }
  \label{fig:tts-pipe}
\end{figure}

Figure~\ref{fig:tts-pipe} provides an overview of the TTS pipeline used in our experiments. 
The system follows the standard mel-spectrogram-based architecture and consists of the following stages:
\[
  \text{Arabic text}
  \;\xrightarrow{\;\text{phonemizer}\;}\;
  \text{phonemes}
  \;\xrightarrow{\;\text{tokenizer}\;}\;
  \text{token IDs}
  \;\xrightarrow{\;\text{acoustic model}\;}\;
  S_{\text{syn}}(b,m)
  \;\xrightarrow{\;\text{vocoder}\;}\;
  s_{\text{syn}}(n).
\]

The input Arabic sentence is first converted into a phonemic sequence by a \emph{phonemizer}. 
Each phoneme is then mapped to a unique token ID by a \emph{tokenizer}. 
Speech generation is split into two stages: the \emph{acoustic model} predicts a log-mel spectrogram from token IDs, and the \emph{vocoder} synthesizes a waveform from the generated spectrogram.

\begin{itemize}
  \item \textbf{Phonemization.}
  Applying a phonemizer before tokenization is common when pronunciation rules are hard to learn from text alone or when explicit phoneme-level control is desirable. 
  For fully \emph{diacritized} Modern Standard Arabic (MSA), the grapheme--phoneme mapping is relatively deterministic, and large datasets allow models to learn these rules directly.
  
  \item \textbf{Unvowelized text.}
  In practical scenarios most Arabic text is unvowelized.
  Automatic vowelizers or diacritizers~\cite{Barqawi2017,Fadel_2019,alasmary2024cattcharacterbasedarabictashkeel} 
  have seen substantial advances in recent years and can reliably recover a large portion of missing diacritics in Modern Standard Arabic. 
  However, they are still primarily optimized for MSA and may introduce occasional errors, which can propagate into the TTS pipeline if not carefully post-processed.
  Training a robust TTS system on unvowelized input typically requires substantially larger datasets than the ASC, for which reliable reference corpora are currently lacking. 
  As automatic speech recognition (ASR) accuracy improves, collecting large-scale training corpora through automatic transcription is becoming a promising strategy.

  \item \textbf{Two-stage vs.\ one-stage architectures.}
  The two-stage design offers several advantages:
  (i) token-to-mel models train quickly and are stable;
  (ii) vocoders \cite{kong2020hifigangenerativeadversarialnetworks,siuzdak2024vocosclosinggaptimedomain} can be trained independently, even in a self-supervised manner; and
  (iii) different acoustic models can reuse the same vocoder, simplifying experimentation.  
  One-stage models such as VITS~\cite{kim2021vits} eliminate the mel-spectrogram bottleneck and can achieve high naturalness, but they require slower training, heavier adversarial objectives, and generally larger datasets to converge reliably.  
  The choice between a two-stage and one-stage system therefore depends on the target application and available computational resources.
\end{itemize}

\subsection{Audio Processing} \label{sec:audio}

\paragraph{Preprocessing}

For all experiments, audio waveforms are resampled to a sampling rate of 
22{,}050~Hz. Silence segments longer than 200~ms are trimmed. To reduce 
low-frequency noise, spectral components below 60~Hz are removed. Finally, 
all signals are normalized to a target level of $-22$~dBFS.

\paragraph{Log-Mel Spectrogram Extraction}

Acoustic features are represented as log-mel spectrograms. 
The used parameters were first introduced in the HiFi-GAN vocoder~\cite{kong2020hifigangenerativeadversarialnetworks} model.
Given an input waveform $x(n)$ sampled at 22{,}050~Hz, 
a short-time Fourier transform (STFT) is computed with 
an FFT size of $n_{\text{fft}}=1024$, a window length of 
$w=1024$ samples, and a hop size of $h=256$ samples, 
corresponding to a frame rate of 
$\tfrac{\text{sample rate}}{h} \approx 86$~frames/s. 
Each frame is multiplied with a \emph{Hann window} before the transform.

The magnitude spectrum is projected onto a mel filterbank with 
$n_{\text{mels}}=80$ bands, covering the frequency range from 
$f_{\min}=0$~Hz to $f_{\max}=8000$~Hz. The filterbank follows 
the \emph{Slaney-style} mel scaling~\cite{slaney1998auditory}, 
which normalizes filter weights to equalize energy across bands. 
The result is then stabilized by clamping to a minimum of $10^{-5}$ 
and converted to log amplitude:
\[
  S_{\text{mel}}(b,m) = \ln \Big( \max\big( \text{Mel} \cdot |X(k,m)| , 10^{-5}\big) \Big),
\]
where $b$ is the mel band index, $m$ is the frame index, and 
$X(k,m)$ denotes the STFT with DFT bin index $k$.

Reflection padding of $\tfrac{n_{\text{fft}}-h}{2}=384$ samples 
is applied at the signal boundaries, and frames are extracted without 
internal centering (\texttt{center=False}).

\paragraph{Pitch Extraction}
FastPitch optionally conditions the acoustic decoder on framewise $f_{0}$ values. 
Although not strictly required, providing explicit pitch information typically improves prosody modeling and also enables controllable pitch manipulation at inference time. 
We extract the pitch contours using the \emph{pYIN} algorithm~\cite{mauch2014pyin}, computed with the same hop size as the mel-spectrograms to ensure frame-level alignment.

\subsection{Text Processing} \label{sec:text}

The Arabic input text is first converted to a phonemic representation. 
Phonemization is performed using a simplified version of Nawar Halabi's \emph{Arabic Phonetiser}\footnote{Available at \url{https://github.com/nawarhalabi/Arabic-Phonetiser}}, which applies \emph{Buckwalter transliteration} as an intermediate step. 
In the simplified variant used here, emphatic vowels are not assigned distinct symbols, and other context-dependent vowel distinctions are omitted, as such phonetic contexts are expected to be learned implicitly by the acoustic model. 
Each phoneme is then mapped to a unique token ID, with geminated (doubled) consonants represented by the consonant followed by a dedicated doubling token.

\subsection{Models} \label{sec:models}

For the acoustic model we adopt \emph{FastPitch}, chosen for its open-source availability, fast and stable training dynamics, and efficient inference. 
Its modular architecture also makes it easy to extend or modify individual components.  
Since FastPitch predicts log-mel spectrograms rather than waveforms, a neural vocoder is used to synthesize the final audio signal.

\subsubsection{FastPitch} \label{sec:models-fp}

FastPitch~\cite{Lancucki2020FastpitchPT} is a \emph{non-autoregressive} 
text-to-mel model derived from FastSpeech~\cite{ren2019fastspeech}. 
Our implementation is based on NVIDIA's \emph{FastPitch 1.1 for PyTorch}\footnote{Available at \url{https://github.com/NVIDIA/DeepLearningExamples/}}.

An encoder maps the input sequence of $L$ tokens into $L$ latent vectors 
using a stack of \emph{feed-forward Transformer} (FFTr) blocks with self-attention. 
Based on these latents, token-level \emph{durations}, \emph{pitch}, and \emph{energy} are predicted. 
The input sequence is then expanded according to predicted durations, and 
augmented with pitch and energy values before being passed to a parallel 
decoder, which generates the log-mel spectrogram in a single forward pass.

To provide reliable supervision for the \emph{duration predictor}, FastPitch uses a lightweight \emph{alignment network}. This module takes token embeddings and the reference spectrogram as input, and applies cross-attention to compute soft alignments. It is trained with the \emph{alignment loss} introduced in 
RadTTS~\cite{Shih2021RADTTSPF,badlani2021ttsalignmentrule}, combined with a 
\emph{beta-binomial prior} to encourage monotonic, diagonal alignments. 
The resulting alignments yield duration targets for the explicit duration 
predictor.

FastPitch thereby offers explicit prosodic control through pitch and duration, 
while retaining the efficiency of non-autoregressive generation.

\subsubsection{Vocoder: HiFi-GAN} \label{sec:models-voc}

We use HiFi-GAN~\cite{kong2020hifigangenerativeadversarialnetworks} as the vocoder to convert predicted log-mel spectrograms into time-domain waveforms. 
HiFi-GAN is a lightweight, GAN-based neural vocoder that achieves high audio quality while maintaining real-time or faster-than-real-time generation.  

Starting from the official universal HiFi-GAN checkpoint, we fine-tuned the model on the ASC training split for a few hundred iterations to better match the spectral characteristics of the corpus. 
The resulting checkpoint is available in the accompanying repository.\footnote{Available at \url{https://github.com/nipponjo/tts-arabic-pytorch}}
In our experiments, we employ a bias-denoising strength of~0.003.


\subsection{Loss Functions} \label{sec:losses}

Loss design plays a central role in neural TTS, as it determines which aspects of the acoustic signal the model prioritizes during training. 
Standard regression losses such as L1 or L2 encourage predictions that minimize pointwise errors in the (log) mel-spectrogram, but they also bias the model toward smooth, averaged outputs.
Augmenting the objective with additional terms, such as duration, pitch, and energy losses, shapes how the model interpolates between training examples and constrains prosodic variation. 
The choice of loss therefore determines which types of deviations are more acceptable (e.g., tolerating phase or fine-detail errors in exchange for stable magnitudes) and strongly influences the perceptual quality of the synthesized speech.

\subsubsection{FastPitch Default Loss}

The default FastPitch objective is a weighted sum of several loss terms. 
For spectrogram reconstruction, the model minimizes an L2 loss $L_{\text{mel}}$ on the predicted log-mel frames. 

The \emph{duration predictor} is trained using
\[
  L_{\text{dur}}
  = \bigl|\log(\hat{d}+1) - \log(d+1)\bigr|^2,
\]
where $\hat{d}$ denotes the predicted frame durations and $d$ the durations produced by the alignment network. 
The \emph{alignment network} itself is trained in a self-supervised manner from the mel-spectrogram and token embeddings using the loss described in \cite{Shih2021RADTTSPF,badlani2021ttsalignmentrule}.  
A binarization term $L_{\text{bin}}$ encourages the soft alignment matrix to approach a hard, monotonic path.

Pitch is normalized to zero mean and unit variance using global statistics, with unvoiced frames set to~zero. 
The \emph{pitch predictor} is optimized via an MSE loss $L_{\text{pitch}}$ between predicted and normalized ground-truth pitch values.

The model also includes an \emph{energy predictor}, where frame-level energy is defined as the L2 norm of each log-mel frame.  
The corresponding loss $L_{\text{energy}}$ is again an MSE term between predicted and target energy sequences.

The full default training objective is therefore
\begin{align} \label{eq:loss-def}
  L_{\text{default}}
  = L_{\text{mel}}
  + L_{\text{dur}}
  + L_{\text{pitch}}
  + 0.1\,L_{\text{energy}}
  + L_{\text{align}}
  + L_{\text{bin}}.
\end{align}

\subsubsection{Standard Reconstruction Loss} \label{sec:loss-standard}

\paragraph{Why $L_p$ Losses Favor Averages}

When training with standard $L_p$ losses, the model is encouraged to make
\emph{average} predictions whenever the training data exhibit variability
that it cannot predict. With an $L_2$ loss, the model learns to predict
the mean of the possible outcomes, while with an $L_1$ loss it predicts
the median \cite{hastie2009elements}. In both cases, this has the practical effect that fine but unpredictable details are smoothed out, because the model replaces them with a central tendency.
This explains why $L_p$-based training often
produces oversmoothed spectrograms in TTS: variation in pitch or
high-frequency structure that is difficult to model is averaged away,
leading to less natural results.

\paragraph{How $L_2$ Attenuates Unpredictable Detail}
The averaging effect of $L_2$ can be analyzed in the spectral domain. Let
$\hat{\mathbf x},\mathbf x\in\mathbb{R}^N$ denote a model prediction and a
reference signal, and let $\mathbf F$ be the \emph{unitary} DFT matrix
($\mathbf F^{-1}=\mathbf F^{\mathrm H}$). The squared $L_2$ loss is
\begin{align}
  L_2 &= \|\hat{\mathbf x}-\mathbf x\|_2^2
       = (\hat{\mathbf x}-\mathbf x)^{\mathrm H}(\hat{\mathbf x}-\mathbf x).
\end{align}
By Parseval's theorem,
\begin{align}
  L_2 &= \|\hat{\mathbf S}-\mathbf S\|_2^2
       = \big(\hat{\mathbf S}-\mathbf S\big)^{\mathrm H}\big(\hat{\mathbf S}-\mathbf S\big),
  \quad \hat{\mathbf S}=\mathbf F\hat{\mathbf x},\;\mathbf S=\mathbf F\mathbf x .
\end{align}
Writing each spectral bin in polar coordinates, $S_k=|S_k| e^{j\varphi_k}$
and $\hat S_k=|\hat S_k| e^{j\hat\varphi_k}$, yields the binwise contribution
\begin{align}
  |\hat S_k-S_k|^2
  = |\hat S_k|^2 + |S_k|^2 - 2\,|\hat S_k|\,|S_k|\cos(\hat\varphi_k-\varphi_k).
\end{align}
The gradients w.r.t.\ predicted magnitude and phase are
\begin{align}
  \frac{\partial L_2}{\partial |\hat S_k|}
    &= 2|\hat S_k| - 2|S_k|\cos(\Delta\varphi_k),\\
  \frac{\partial L_2}{\partial \hat\varphi_k}
    &= 2\,|\hat S_k|\,|S_k|\,\sin(\Delta\varphi_k),
  \qquad \Delta\varphi_k=\hat\varphi_k-\varphi_k .
\end{align}
If the phase error $\Delta\varphi_k$ is essentially random (e.g., uniformly
distributed), then
$\mathbb E[\cos(\Delta\varphi_k)]=\mathbb E[\sin(\Delta\varphi_k)]=0$, giving
\begin{align}
  \mathbb E\!\left[\tfrac{\partial L_2}{\partial |\hat S_k|}\right] = 2|\hat S_k|,
  \qquad
  \mathbb E\!\left[\tfrac{\partial L_2}{\partial \hat\varphi_k}\right] = 0 .
\end{align}

Thus the expected negative gradient drives $|\hat S_k|$ toward zero, while no
systematic phase update occurs. 
Intuitively, whenever the phase is
unpredictable, the $L_2$ loss cannot exploit the cross-term and instead shrinks
the magnitude. This explains why models trained with $L_2$ tend to suppress
spectral bins with random phase—typically high-frequency components—and
thereby produce smoother, less detailed spectra.
 
Consider a frequency bin where the training data has a
consistent magnitude but random phase across utterances. From the model's
perspective, the phase is unpredictable. Under $L_2$, the expected gradient
reduces the predicted magnitude in that bin, effectively silencing it.

\subsubsection{Adversarial Loss} \label{sec:loss-adv}
To encourage the generation of more natural mel-spectrograms, we incorporate 
an additional adversarial loss and analyze its impact in our experiments.

\paragraph{Discriminator}
The discriminator design follows the lightweight convolutional discriminator 
employed in the image inpainting model \textit{DeepFillv2}~\cite{yu2019freeformimageinpaintinggated}, adapted here to operate on log-mel spectrograms in order to reduce oversmoothing in TTS training.
The discriminator is composed of five $5{\times}5$ two-dimensional convolutional 
layers with stride $2$. Each layer is followed by a LeakyReLU activation 
($\alpha=0.2$), and spectral normalization~\cite{miyato2018spectralnormalization} is applied to all convolutional 
weights to improve stability. The input consists of spectrogram segments of 
128 consecutive frames, randomly cropped from the batch of full log-mel 
spectrograms. 

In the multi-speaker setting, the discriminator is speaker-conditioned. The 
speaker embedding is projected through two fully connected layers, each with 
spectral normalization and LeakyReLU activation ($\alpha=0.2$), to produce a 
vector of length equal to the number of mel channels. This vector is repeated 
across the time axis and concatenated as an additional input channel to the 
spectrogram.

\paragraph{Loss Functions}
We adopt the \emph{least-squares GAN} (LS-GAN) formulation~\cite{mao2017squaresgenerativeadversarialnetworks} 
for the adversarial objective. The discriminator $D$ is trained to distinguish 
natural log-mel spectrograms $S_{\text{ref}}$ from generated ones $S_{\text{pred}}$, while 
the generator $G$ aims to minimize both reconstruction losses and the adversarial loss:
\begin{align}
  L_D &= \tfrac{1}{2}\big(D(S_{\text{ref}})-1\big)^2 
       + \tfrac{1}{2}\big(D(S_{\text{pred}})\big)^2, \\
  L_G &= \big(D(S_{\text{pred}})-1\big)^2.
\end{align}
In addition, a \textit{feature matching loss} is employed, defined as the mean 
absolute error (MAE) between the feature maps of the first four discriminator 
layers for reference and synthesized spectrograms.

During training, the generator loss $L_G$ is combined with the model's 
default loss $L_{\text{default}}$ to form the overall 
objective
\begin{equation} \label{eq:loss-fp-adv}
    L = L_{\text{default}} + \alpha L_G.
\end{equation}
In preliminary experiments we found $\alpha = 4$ to provide a good balance 
between reconstruction fidelity and adversarial regularization, although a 
more extensive hyperparameter study remains for future work.

\subsection{Oversmoothing Metrics} \label{sec:eval-smooth}

\begin{figure}[H]
  \centering
  \includegraphics[width=0.75\linewidth]{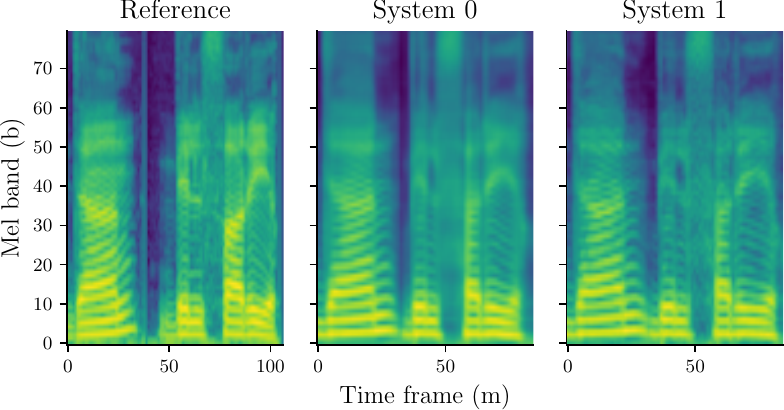}
  \caption{Log-mel spectrograms of a reference utterance and outputs from two 
  TTS systems. System~0 exhibits smoother spectral patterns with reduced 
  high-frequency detail compared to the reference, whereas System~1 preserves 
  more fine structure. The vertical axis shows mel frequency bands ($b$), and 
  the horizontal axis shows time frames ($m$).}
  \label{fig:specs_ref_s0_s1}
\end{figure}

\paragraph{Motivation}
Oversmoothing refers to the tendency of neural TTS models to produce
spectrograms with reduced variance and overly smooth trajectories.
This phenomenon is commonly attributed to the use of simple pointwise
$L_{p}$ reconstruction losses on (log) mel-spectrograms, which encourage
average predictions and suppress high-frequency detail.

A common way to analyze oversmoothing is to examine the spectral 
variability across the mel-bin axis. This can be achieved by applying 
a real-input FFT (rFFT)\footnote{We follow the convention of common libraries (e.g., NumPy, SciPy) 
where \texttt{rFFT} denotes the FFT for real-valued inputs. The transform 
itself is still complex-valued; the optimization comes from discarding the 
redundant negative-frequency components implied by Hermitian symmetry.} 
along the mel dimension of each frame, yielding what we may call a \emph{mel-cepstrogram}. This representation highlights local variations between adjacent mel channels, thereby exposing the degree of fine-grained spectral detail that may be suppressed by oversmoothing.

\paragraph{Mel-Cepstrogram}
Formally, given a log-mel spectrogram $S(b,m)$ with mel-bin index $b$ 
and frame index $m$, the transform is defined as
\begin{align}
  C_{\text{mel}}(q,m)
    &= \operatorname{rFFT}_{b}\!\big(S^\blacktriangle (b,m)\big),
    \label{eq:mel-cep}
\end{align}
where $q$ denotes the \emph{mel-quefrency index}. 
Before the rFFT, we \emph{subtract the framewise mean} and apply a \emph{Hann window} to each frame, yielding $S^\blacktriangle (b,m)$. 
These preprocessing steps mitigate spectral leakage caused by discontinuities at the mel-band edges. 
The resulting cepstral coefficients emphasize \emph{relative spectral structure} rather than global energy offsets.

\begin{itemize}
  \item \emph{Low values} of $q$ capture \emph{slowly varying structure} across mel bins (broad spectral envelopes), while \emph{higher $q$ indices} reflect \emph{rapid variations} corresponding to \emph{fine spectral detail}, including harmonic detail and noise-like components. 
  \item Since the model outputs log-mel spectrograms, the mel-cepstrogram is 
  computed directly in the log domain. This ensures consistency with the 
  representation used during training, with standard visualization 
  practices, and with the approximately logarithmic sensitivity of human 
  loudness perception.
  \item This procedure is closely related to the standard 
  \emph{mel-frequency cepstral coefficients} (MFCCs), which apply a 
  discrete cosine transform (DCT) to the log-mel energies. 
  Our use of the rFFT across mel bins differs mainly in the 
  choice of transform basis (Fourier vs.\ cosine): while both 
  serve to decorrelate adjacent mel bands and compactly represent 
  spectral detail, the rFFT is particularly suitable here because 
  it yields a frequency-domain representation whose squared 
  magnitude can be directly interpreted as a power distribution 
  over quefrency.
\end{itemize}

\paragraph{Metric Definitions}
To quantify oversmoothing during training and for model comparison, 
we define a set of scores derived from the mel-cepstrogram. 
Excessive smoothing in the synthesized spectrogram manifests as a \emph{reduction in energy at higher $q$ values}, indicating a loss of spectral richness.

Based on this intuition, we define four complementary scores---\emph{High-Quefrency Energy Ratio (HQER)}, \emph{Cepstral Slope}, \emph{Cepstral Centroid}, and \emph{Cepstral Rolloff}---which each quantify different aspects of the high-quefrency energy distribution.
\\\\
Let $P(q,m)$ denote the \emph{power} at \emph{quefrency index} $q$ and \emph{frame index} $m$:  
\begin{align}
  P(q,m) &= \big| C_{\text{mel}}(q,m) \big|^{2}, 
  \qquad q = 0, \dots, Q-1,
\end{align}
where $C_{\text{mel}}(q,m)$ denotes the complex mel-cepstrogram, and 
$Q = \lfloor B/2 \rfloor + 1$, where $B$ is the number of mel bands.

Since only the magnitude is retained, the \emph{phase} of $C_{\text{mel}}$ is 
discarded. This choice is reasonable in the context of analyzing 
oversmoothing, as high-quefrency details primarily reflect fine 
spectral variations often caused by turbulent airflow, whose phase 
structure is largely chaotic and perceptually less important.\footnote{This treatment is consistent with conventional cepstral analysis, where 
only the log-magnitude is retained and phase information is 
commonly ignored, as it contributes little to the perceived timbre 
\cite{rabiner1978digital}.}

\begin{figure}[!htbp]
  \centering
  \begin{subfigure}[t]{0.48\textwidth}
    \centering
    \includegraphics[width=\linewidth]{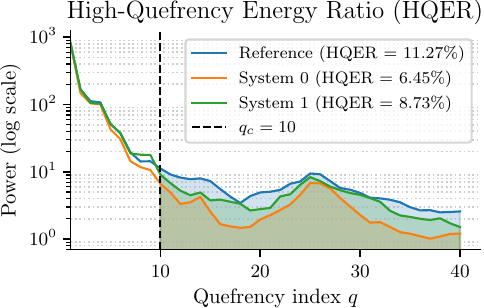}
    \caption{\textbf{High-Quefrency Energy Ratio (HQER)}. Shaded region illustrates high-quefrency energy beyond cutoff $q_c=10$ (log-scale, so area is not proportional).}
    \label{fig:subfig1}
  \end{subfigure}
  \hfill
  \begin{subfigure}[t]{0.48\textwidth}
    \centering
    \includegraphics[width=\linewidth]{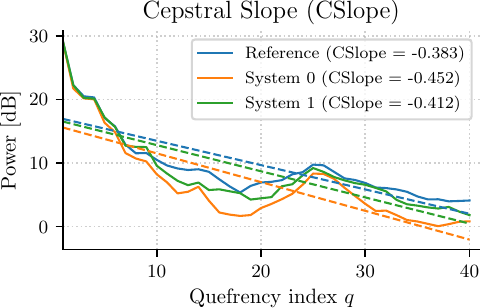}
    \caption{\textbf{Cepstral Slope (CSlope)}. Regression slope of log-power vs. quefrency, reflecting decay rate of fine spectral detail.}
    \label{fig:subfig2}
  \end{subfigure} 

  \par\medskip
  \begin{subfigure}[t]{0.48\textwidth}
    \centering
    \includegraphics[width=\linewidth]{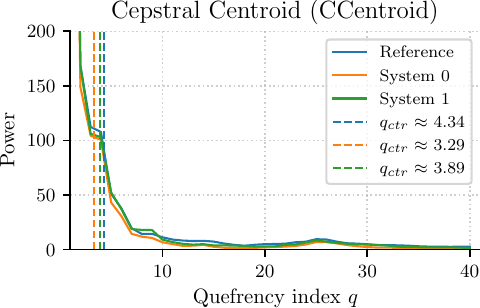}
    \caption{\textbf{Cepstral Centroid (CCentroid)}. Energy-weighted average quefrency index $q_{\text{ctr}}$, lower values indicate smoother spectra.}
    \label{fig:subfig3}
  \end{subfigure}
  \hfill
  \begin{subfigure}[t]{0.48\textwidth}
    \centering
    \includegraphics[width=\linewidth]{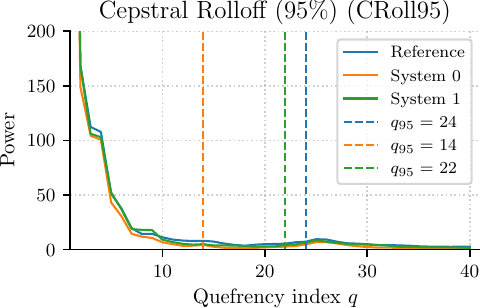}
    \caption{\textbf{Cepstral Rolloff (95\%) (CRoll95)}. Quefrency index $q_{95}$ below which 95\% of cepstral energy is contained.}
    \label{fig:subfig4}
  \end{subfigure}

  \caption{Visualization of cepstral-domain oversmoothing metrics for the reference and the two TTS systems from Figure~\ref{fig:specs_ref_s0_s1}. 
(a) High-Quefrency Energy Ratio (HQER) quantifies energy beyond a cutoff $q_c=10$. 
(b) Cepstral Slope (CSlope) measures the decay of high-quefrency energy. 
(c) Cepstral Centroid (CCentroid) gives the energy-weighted center of cepstral mass. 
(d) Cepstral Rolloff (95\%) (CRoll95) indicates the quefrency index covering 95\% of total cepstral energy. For visualization, curves are averaged across multiple frames.}
  \label{fig:os_metrics_viz}
\end{figure}

Figure~\ref{fig:os_metrics_viz} illustrates the metrics for the example in Figure~\ref{fig:specs_ref_s0_s1}.

\paragraph{High-Quefrency Energy Ratio (HQER)}
\begin{align}
  \text{HQER}(m) &= 
    \frac{\sum_{q \ge q_c} P(q,m)}{\sum_{q=1}^{Q-1} P(q,m)},
\end{align}
where $q_c$ is a cutoff index (e.g., $q_c=\lfloor 0.25Q \rfloor$). 
HQER measures the \emph{proportion of energy contained in the high-quefrency 
region}. Lower values indicate stronger oversmoothing, as less energy 
remains beyond the cutoff. For convenience, HQER can be reported in 
percent by multiplying by $100$.
An analogy is the fraction of ``weight'' found in the tail of a 
distribution: if little mass lies beyond the cutoff, the signal is 
dominated by coarse structure.

\paragraph{Cepstral Slope (CSlope)}
\begin{align}
  \text{CSlope}(m) &= 
    \operatorname{linreg}_{q=1,\dots,Q-1}\!
    \Big(q, 10\log_{10}\big(P(q,m)+\varepsilon\big)\Big),
\end{align}
defined as the least-squares \emph{slope of the log-power spectrum (in dB)} 
with respect to quefrency $q$. A more negative slope indicates stronger 
low-pass characteristics and thus more smoothing. The unit is 
decibels per quefrency bin (dB/bin).
Geometrically, CSlope is like the tilt of a ramp: a steep downward ramp 
means that fine detail (high-$q$ energy) decays quickly.

\paragraph{Cepstral Centroid (CCentroid)}
\begin{align}
  \text{CCentroid}(m) &= 
    \frac{\sum_{q=1}^{Q-1} q \, P(q,m)}{\sum_{q=1}^{Q-1} P(q,m)},
\end{align}
the \emph{energy-weighted mean quefrency index}. This is directly analogous to the \emph{center of mass}, where 
$P(q,m)$ acts as the mass distribution across quefrency indices. Lower 
centroid values indicate that the ``mass'' of the cepstral energy is 
concentrated at low quefrencies, reflecting smoother spectral shapes.

\paragraph{Cepstral 95\% Rolloff (CRoll95)}
\begin{align}
  \text{CRoll95}(m) &= \min \Big\{ q : 
    \tfrac{\sum_{r=1}^{q} P(r,m)}{\sum_{r=1}^{Q-1} P(r,m)} \ge 0.95 \Big\},
\end{align}
the \emph{smallest quefrency index that accumulates at least $95\%$ of the 
total cepstral energy}. Smaller rolloff values indicate that most energy 
is concentrated at lower quefrencies, again signaling stronger smoothing.
By analogy, CRoll95 is like a \emph{quantile} in statistics: it marks 
the quefrency location below which the bulk of the energy ``mass'' resides.
\\\\
The first three metrics—HQER, Cepstral Slope, and Cepstral Centroid—are differentiable once a small constant~$\varepsilon$ is added to avoid division by zero in denominators. 
The Cepstral Rolloff, while not inherently differentiable, can be approximated using a smooth surrogate function.\footnote{One example is a 
soft quantile approximation: 
\[
  \widetilde{\text{CRoll95}}(m) 
    = \frac{\sum_{q=1}^{Q-1} q \, \softmax\!\big(-\tau |F(q,m)-0.95|\big)}
           {\sum_{q=1}^{Q-1} \softmax\!\big(-\tau |F(q,m)-0.95|\big)},
\]
where $F(q,m)$ is the cumulative energy distribution, 
$\tau$ controls sharpness (larger $\tau$ gives a closer 
approximation to the hard cutoff), and $\softmax$ acts over $q$. This formulation is only one possible differentiable approximation; 
other smooth quantile operators could equally be applied.}
Consequently, all four metrics, as well as the power measure $P(q,m)$, could in principle be integrated into the training objective to directly penalize oversmoothing. 
In this work, however, we employ them solely as evaluation metrics and leave their use in training for future exploration.

\newpage

\subsection{Evaluation Protocol} \label{sec:eval}

During training, we track a set of standard reconstruction metrics as well as the proposed oversmoothing measures using the \emph{ASC test set} described in Section~\ref{sec:data-asc}. All models are evaluated using deterministic inference with predicted durations, energy, and pitch, and without any post-filtering.

For reconstruction quality, we compute the average L1 (MAE) and L2 (MSE) errors and the spectral convergence (SConv)~\cite{Arik_2019} between predicted and reference mel-spectrograms. 
To account for temporal misalignment, we align the predicted spectrograms to the reference using \emph{dynamic time warping} (DTW)~\cite{Sakoe1978DynamicPA, Müller2007} with a framewise cosine distance.

The \emph{speaking rate} (SPR) is measured as the number of tokens produced per second.

For pitch evaluation, $f_{0}$ contours are extracted from the synthesized and reference waveforms using \emph{Praat}~\cite{boersma2024praat} due to its efficient pitch estimator. 
The contours are DTW-aligned using the L2 distance, after which we compute 
the root mean squared error (RMSE), the Pearson correlation coefficient~$r$, and the voiced/unvoiced (V/UV) error rate.

For oversmoothing, we report the framewise MAE of each cepstral-domain metric after DTW alignment 
(again using the L2 distance), allowing direct comparison of local spectral-contrast differences.

\paragraph{Utterance-level Metrics}
In addition to framewise measures, we also track differences in 
\emph{utterance-level statistics} of the pitch:
\begin{align}
  \Delta_u \mu f_0 
    &= \mu(\hat{f}_0) - \mu(f_0), \\
  \Delta_u \sigma f_0 
    &= \sigma(\hat{f}_0) - \sigma(f_0),
\end{align}
where $\mu(\cdot)$ and $\sigma(\cdot)$ denote the mean and standard deviation of the voiced frames within the utterance, $\hat{f}_0$ and $f_0$ denote framewise pitch contours for the synthesized and reference utterance.
The mean difference $\Delta_u \mu f_0$ captures shifts in overall pitch height, 
whereas the standard deviation difference $\Delta_u \sigma f_0$ reflects changes in pitch variability and thus perceived expressiveness.

We compute analogous utterance-level differences for the speaking rate, indicating whether a generated utterance tends to be faster or slower than the reference, and for each oversmoothing metric, indicating whether the model exhibits a global tendency toward over- or undersmoothing.


\subsection{Training Setup} \label{sec:training}

We train all models using the \emph{AdamW} optimizer~\cite{loshchilov2019adamw} with a \emph{weight decay} of $10^{-6}$. 
For the FastPitch baseline, we use a \emph{learning rate} of $10^{-4}$, no gradient clipping, and \emph{momentum} parameters $(\beta_{1}, \beta_{2}) = (0.9, 0.999)$.

When applying \emph{adversarial training}, both the generator and discriminator are optimized with $(\beta_{1}, \beta_{2}) = (0.0, 0.99)$. 
We observed that using a positive $\beta_{1}$ led to periodic oscillations in the adversarial loss curves, whereas setting $\beta_{1}=0$ resulted in smooth and stable training dynamics.

All models are trained for 200k batches. 
In the single-speaker setup, one epoch consists of 198 batches, while in the four-speaker configuration an epoch comprises 792 batches. 
Evaluation metrics are computed at the end of each epoch, and checkpoints are saved every 1000 steps.


\pagebreak

\section{Results} \label{sec:results}

We report results for three experimental configurations designed to assess the effects of training objectives and speaker diversity on Arabic FastPitch performance. 
First, we establish a baseline using the original single-speaker ASC dataset and the standard regression loss described in Eq.~\ref{eq:loss-def} (Section~\ref{sec:res-std}). 
Next, we investigate how augmenting the training objective with an adversarial spectrogram loss (Eq.~\ref{eq:loss-fp-adv}) affects reconstruction accuracy, pitch modeling, and spectral oversmoothing (Section~\ref{sec:res-adv}). 
Finally, we evaluate a multi-speaker model trained with three additional synthetic speakers generated using XTTSv2 under the same adversarial objective (Section~\ref{sec:res-multispk}). 

\subsection{Baseline with Default Loss} \label{sec:res-std}

\paragraph{Quantitative Evaluation}
Table~\ref{tab:fpstd-metrics} summarizes the evolution of objective metrics for the FastPitch model trained with the default $L_2$ loss. All measures show consistent improvement from early training (epoch~1) to epoch~1000, with most reaching their optimal values between epochs~40 and~300. 
The mel-spectrogram distances (L1, L2, SConv) decrease rapidly during the first 50~epochs and then plateau, indicating fast convergence of the decoder. 
Pitch-related metrics follow a similar trend, with correlation~$r$ improving steadily and the $\Delta f_{0}$-RMSE decreasing to approximately~7~Hz. 
The cepstral oversmoothing metrics (HQER, CSlope, CCentroid, CRoll95) contiue to decrease for a longer time, indicating that fine-grained spectral contrast continues to improve even after reconstruction losses have plateaued. 

\begin{table}[h]
\centering
\caption{
Evaluation metrics for the FastPitch model trained with the default ($L_2$) loss. 
For each measure, the first column reports the best validation value and the training epoch at which it occurs. 
The second column shows the metric value after the first training epoch and the mean $\pm$ standard deviation for epochs 970-1000, illustrating convergence behavior over time. 
Lower values indicate better performance for all metrics except the Pearson correlation~$r$, where higher is better.
}
\label{tab:fpstd-metrics}
\begin{tabular}{lll}
\toprule
Measure & Best @ epoch & @1 $\to$ @[970-1000] \\
\midrule
L1 & 0.723 @ 52 & 1.13 $\to$ 0.746 $\pm$ 0.0025 \\
L2 & 0.948 @ 40 & 1.46 $\to$ 0.973 $\pm$ 0.0029 \\
SConv & 0.166 @ 53 & 0.264 $\to$ 0.170 $\pm$ 0.00042 \\
$\Delta f_{0, RMSE}$/Hz & 6.90 @ 256 & 14.1 $\to$ 7.36 $\pm$ 0.078 \\
Pearson $r$ & 0.953 @ 310 & 0.868 $\to$ 0.950 $\pm$ 0.0010 \\
$E_{V/UV}$ & 0.272 @ 241 & 0.358 $\to$ 0.283 $\pm$ 0.0036 \\
MAE(HQER)/\% & 5.49 @ 869 & 13.3 $\to$ 5.82 $\pm$ 0.11 \\
MAE(CSlope)/$\frac{dB}{bin}$ & 0.144 @ 620 & 0.351 $\to$ 0.155 $\pm$ 0.0036 \\
MAE(CCentroid)/bin & 1.36 @ 936 & 3.44 $\to$ 1.41 $\pm$ 0.023 \\
MAE(CRoll95)/bin & 5.39 @ 936 & 16.0 $\to$ 5.62 $\pm$ 0.084 \\
\bottomrule
\end{tabular}
\end{table}

\begin{figure}[H]
  \centering  
  \includegraphics[width=0.55\linewidth]{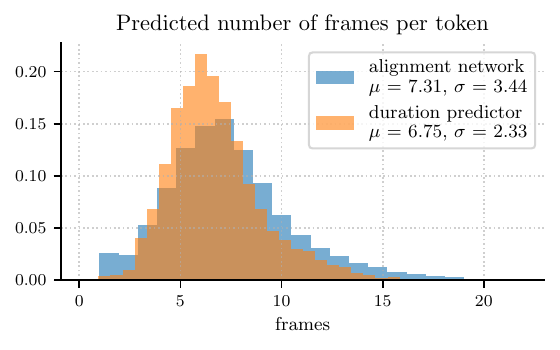}
  \vspace{-3mm}
  \caption{Normalized histograms of predicted frame durations per token for the alignment network (blue) and the learned duration predictor (orange) on the test set. 
    The predictor closely follows the alignment distribution but with a slightly lower mean ($\mu = 6.75$ vs.\ $7.31$) and reduced variance ($\sigma = 2.33$ vs.\ $3.44$), indicating mild compression of longer phoneme durations.}
  \label{fig:duration_hist}
\end{figure}

\paragraph{Duration Prediction}
Figure~\ref{fig:duration_hist} compares the per-token frame durations predicted by the alignment network and the duration predictor. 
The two distributions largely overlap, confirming that the predictor successfully learns the alignment-derived durations used as supervision during training. 
However, the predicted durations exhibits a slightly lower mean and smaller spread than the alignment network, indicating mild compression of longer phoneme durations.

\begin{figure}[H]
  \centering  
  \includegraphics[width=0.55\linewidth]{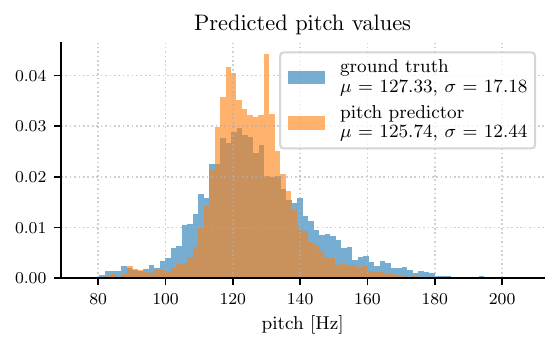}
  \vspace{-3mm}
  \caption{
    Normalized histograms of pitch values predicted by the FastPitch pitch predictor (orange) and ground-truth pitch extracted from the test set (blue). 
    The predictor captures the overall distribution shape but with a slightly lower mean ($\mu = 125.7$~Hz vs.\ $127.3$~Hz) and reduced variance ($\sigma = 12.4$ vs.\ $17.2$), reflecting mild pitch compression typical of regression-based predictors. 
    Note that the test set of the Arabic Speech Corpus exhibits slightly different pitch statistics than the training set (Table~\ref{tab:train_test_metrics}), which contributes to the observed mean shift.
  }
  \label{fig:pitch_hist}
\end{figure}

\paragraph{Pitch Prediction}
Figure~\ref{fig:pitch_hist} compares the predicted pitch values with ground-truth pitch extracted from the test set. 
The model captures the overall shape of the distribution, although the predicted pitch shows a slightly lower mean and reduced variance. 
Part of this shift is attributable to a small systematic difference in pitch statistics between the training and test partitions (Table~\ref{tab:train_test_metrics}).

\begin{figure}[H]
\centering 
\begin{subfigure}{0.9\linewidth}
  \includegraphics[width=\linewidth]{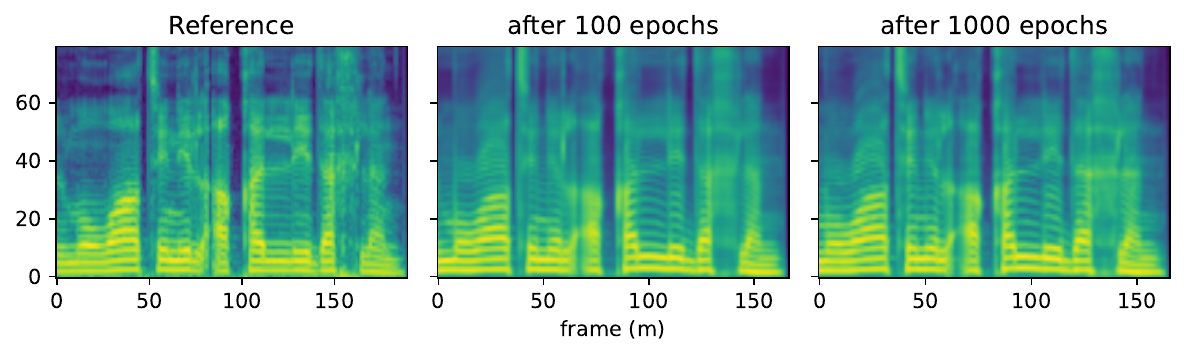}
  \caption{Standard loss}
\end{subfigure}

\vspace{0.5em} 
\begin{subfigure}{0.9\linewidth}
  \includegraphics[width=\linewidth]{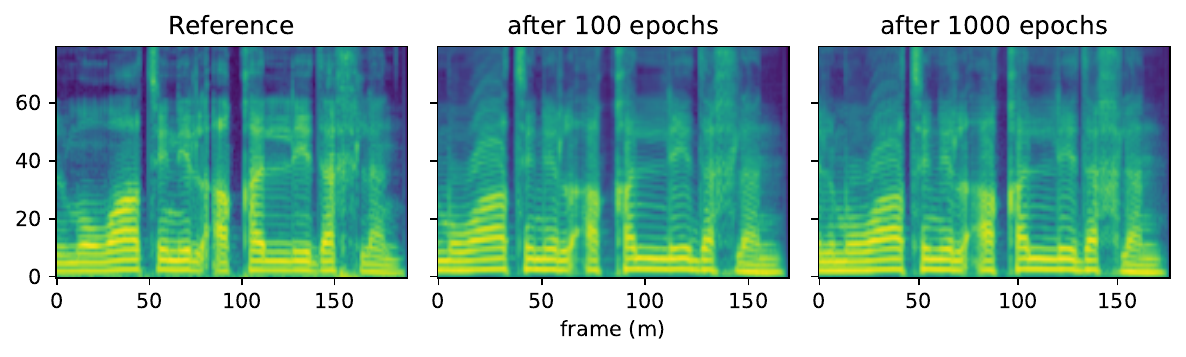}
  \caption{Adversarial loss}
\end{subfigure}

\caption{
Evolution of predicted mel-spectrograms over training for the FastPitch model using (a) standard $L_2$ loss and (b) the proposed adversarial spectrogram loss. 
Each row shows the same utterance reference from the test set (left) and predictions after 100 and 1000 epochs. 
Under the standard loss, spectral details are more blurred, whereas adversarial training preserves higher-frequency detail and sharper spectral contrast, indicating reduced oversmoothing.
}
\label{fig:specs_comp}
\end{figure}

\subsection{Effect of Adversarial Loss} \label{sec:res-adv}

\begin{figure}[H]
  \centering
  \includegraphics[width=0.95\linewidth]{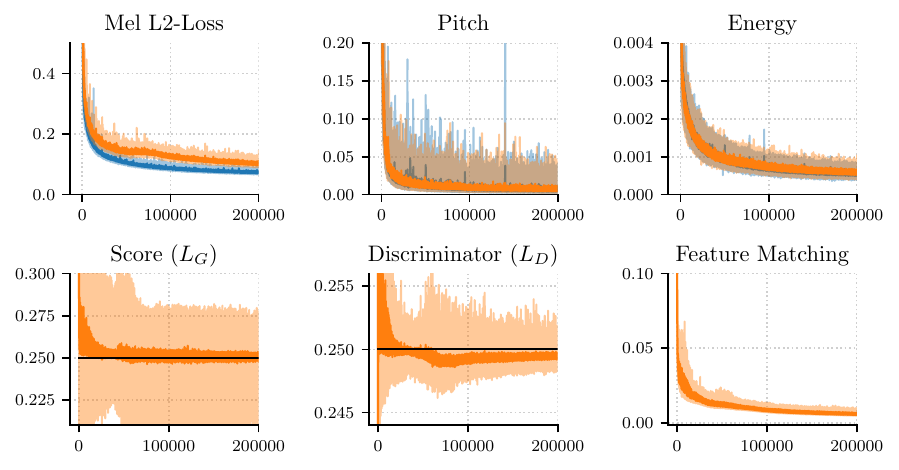}
  \caption{
  Training loss curves for FastPitch under two configurations: the default regression-based objective (blue) and the proposed setup with an additional lightweight adversarial spectrogram loss (orange). 
  The top row shows the main regression losses for mel-spectrogram L2, pitch prediction, and energy prediction. 
  The bottom row displays GAN-related terms, including the generator score $L_G$, discriminator loss $L_D$, and the feature-matching loss. 
  The adversarial model converges smoothly, with $L_G$ and $L_D$ stabilizing near equilibrium values, indicating stable LS-GAN training with spectral normalization. Darker curves show an EMA-smoothed version with a smoothing factor of 0.9. 
  }
  \label{fig:fastpitch_train}
\end{figure}

\begin{figure}[H]
\centering 
\begin{subfigure}{0.9\linewidth}
  \includegraphics[width=\linewidth]{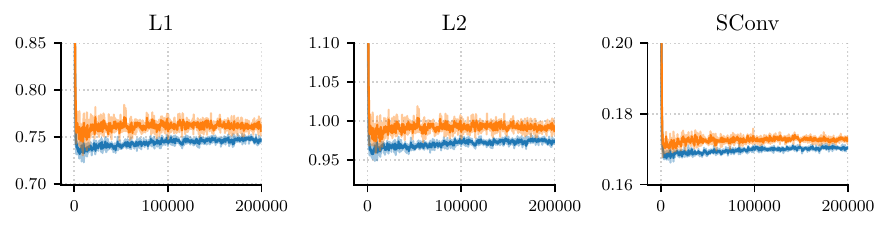}
  \caption{Mel spectrogram distances}
\end{subfigure}

\vspace{0.5em} 
\begin{subfigure}{0.9\linewidth}
  \includegraphics[width=\linewidth]{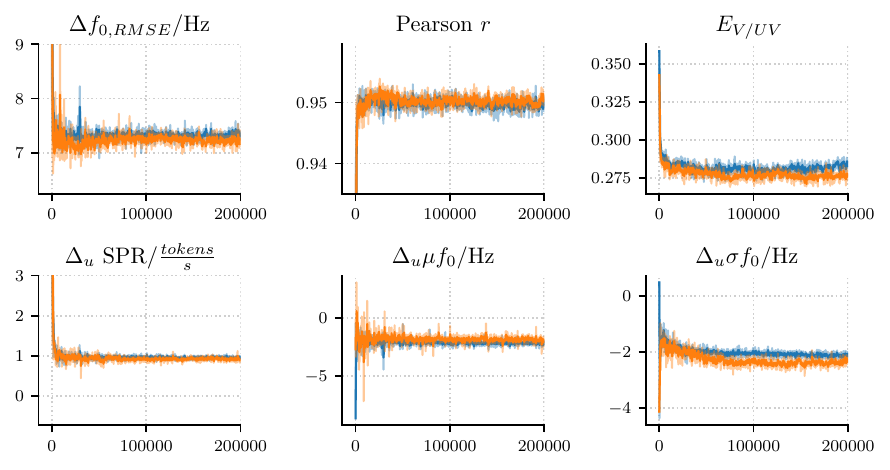}
  \caption{Pitch metrics \& Speaking rate}
\end{subfigure}

\vspace{0.5em}
\begin{subfigure}{0.9\linewidth}
  \includegraphics[width=\linewidth]{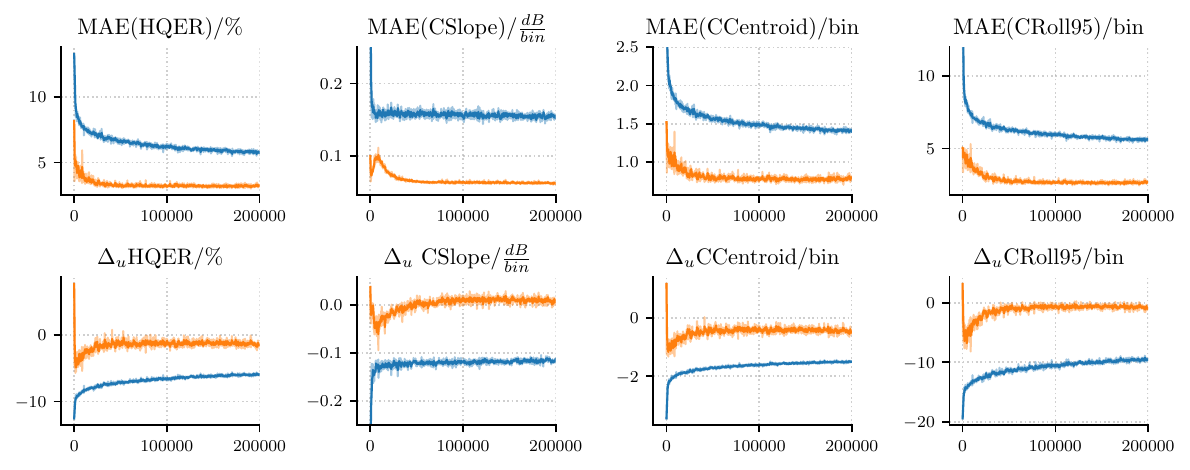}
  \caption{Oversmoothing metrics}
\end{subfigure}

\caption{FastPitch evaluation metric curves with: default loss (blue), additional adversarial loss (orange).}
\label{fig:fastpitch_eval}
\end{figure}

Figures~\ref{fig:fastpitch_train} and~\ref{fig:fastpitch_eval} compare the FastPitch model trained with the default regression objectives (blue) and with an additional adversarial spectrogram loss (orange). The results illustrate the effect of the lightweight LS-GAN with spectral normalization and feature matching on both training dynamics and evaluation metrics.

\paragraph{Training Behavior}
The curves in Figure~\ref{fig:fastpitch_train} show that the adversarially trained model reaches comparable convergence in the core regression objectives (mel L2-loss, pitch, energy, and duration). The discriminator-related losses ($L_D$, feature matching, and generator score $L_G$) remain smooth and bounded, suggesting that the LS-GAN formulation provides stable optimization without mode collapse. Overall, the training remains well-behaved and convergent despite the additional adversarial objective.

\paragraph{Mel-Spectrogram Distances (L1, L2, SConv).}  
The adversarial model shows slightly \emph{higher} L1/L2/SConv distances than the baseline. This is expected, as the adversarial loss encourages the model to reproduce finer but less mean-centered spectral details. These fine-grained variations increase pointwise error even as they improve the naturalness and variability of the generated spectra. Consequently, the marginal rise in reconstruction distance reflects a trade-off between strict sample-wise accuracy and richer spectral texture.

\paragraph{Pitch and Prosodic Metrics}  
Pitch RMSE ($\Delta f_0$) and correlation ($r$) remain comparable between models, with a slight reduction in $f_0$ bias and similar noiced/unvoiced error rate ($E_{V/UV}$). 
The speaking-rate deviation ($\Delta_u \mathrm{SPR}$) and utterance-level pitch statistics ($\Delta_u \mu_{f0}$, $\Delta_u \sigma_{f0}$) are nearly identical, confirming that adversarial training does not negatively impact prosodic timing or stability.

\paragraph{Cepstral Oversmoothing Metrics}  
The most pronounced differences appear in the cepstral-domain measures (HQER, CSlope, CCentroid, CRoll95). 
Both the \emph{framewise mean absolute error} (MAE) and the \emph{utterancewise} ($\Delta_u$) variants show consistently lower values and reduced bias for the adversarial model throughout training. 
For the baseline model trained with the default L2 mel loss, these metrics show a decreasing trend even after 1000 epochs, suggesting gradual transfer of finer spectral detail from training to validation. 
The adversarial model reaches significantly lower cepstral errors early in training and then decreases more slowly compared to the baseline. 
Notably, the cepstral metrics continue to improve even after other validation metrics have begun to plateau or slightly increase.

\subsection{Multi-Speaker Training with Synthetic Data} \label{sec:res-multispk}

\begin{figure}[H]
  \centering
  \includegraphics[width=0.95\linewidth]{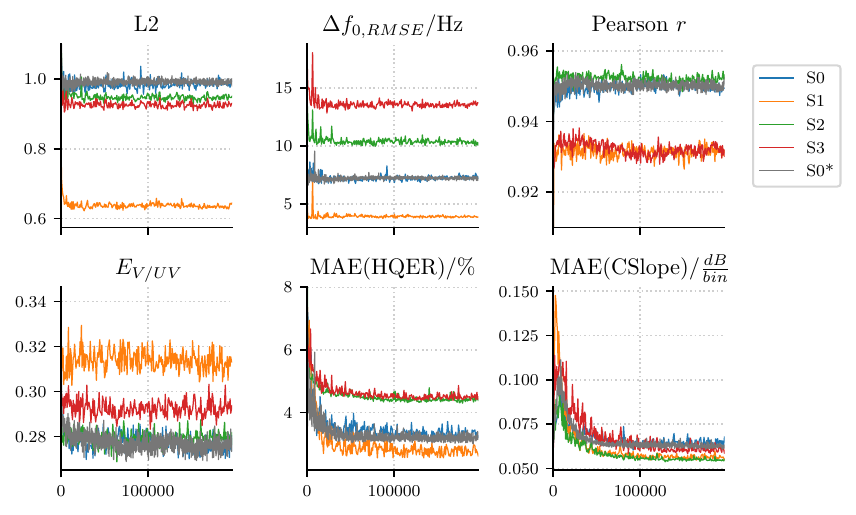}
  \caption{Evaluation metric curves for each speaker S0-S3 and the curves from the single speaker model denoted by S0*. All metrics averaged across all speakers are shown in Figure~\ref{fig:curves-ms-avg}.}
  \label{fig:curves-ms}
\end{figure}

Figure~\ref{fig:curves-ms} shows the evolution of evaluation metrics for the four speakers 
in the multi-speaker FastPitch model: the natural ASC speaker (S0), a synthetic male voice (S1), 
and two synthetic female voices (S2, S3). Metrics from the single-speaker adversarial baseline 
are included as S0$^{*}$ for comparison. Despite large differences in prosodic statistics 
(Table~\ref{tab:speaker_metrics}), the model converges stably for all speakers.

Speakers differ systematically in their pitch-related performance: voices with lower pitch 
variance (e.g., S1) achieve lower $f_{0}$-RMSE throughout training, while speakers with wider 
pitch ranges (S2, S3) show higher pitch errors and slower early convergence but eventually 
stabilize. The synthetic male speaker (S1) also reaches the lowest mel-spectrogram L2 values, 
followed by the synthetic female speakers.

Importantly, performance for the original ASC voice (S0) remains comparable to the single-speaker baseline (S0$^{*}$). Mel L2, pitch RMSE, correlation, V/UV error, and cepstral oversmoothing metrics all remain within the same range, indicating that the addition of three synthetic speakers does not degrade single-speaker quality.

The cepstral oversmoothing metrics (HQER, CSlope, CCentroid) show speaker-dependent offsets, with higher values for S2 and S3. These differences follow the prosodic and spectral properties of 
the speakers and are consistent with their intrinsic pitch characteristics. Across all speakers, the oversmoothing metrics decrease steadily during training and converge to values comparable to the single-speaker model.

Overall, multi-speaker training converges reliably, maintains performance for S0, and models the synthetic speakers effectively.

\section{Discussion} \label{sec:discussion}

The following discussion follows an analogous structure to the results, with each section of the results having its own section.

\subsection{Baseline Training}

The FastPitch model trained stably and quickly learned to generate intelligible Arabic speech from diacritized text converted to phonemes. 

Across all experiments, the synthesized utterances tended to be slightly faster than the corresponding references. 
The duration predictor underestimates long phoneme durations and produces a narrower distribution than the alignment network. 
This compression pattern is characteristic of regression-based duration models, which tend to 
gravitate toward mean values and penalize long-duration outliers more strongly. 
Since slowing down speech without introducing artifacts is more difficult than slightly speeding it up, it may be beneficial in future work to bias the model toward longer durations, for example by using asymmetric loss functions or duration-dependent weighting schemes.

A similar trend is observed for pitch: the predicted pitch distribution exhibits reduced variance relative to the ground truth. 
This reduction in dynamic range follows from the use of $L_2$ regression losses, which disproportionately penalize large deviations and thus dampen high-pitch excursions and rapid local fluctuations. 
Future work could explore pitch-variance-preserving approaches, such as variance-enhancing regularization terms, focal~\cite{lin2018focallossdenseobject} or heteroscedastic losses, or adversarial objectives applied directly to $f_0$ trajectories.

Overall, while the baseline converges smoothly and learns the coarse structure of both duration and pitch, its systematic variance reduction motivates the use of complementary losses or regularization strategies to improve prosodic richness.

\subsection{Adversarial Training}

Adversarial training was found to run stably and robustly under the proposed configuration. 
The discriminator's smooth loss trajectories and the bounded generator scores indicate that the LS-GAN formulation with spectral normalization and feature matching yields stable training dynamics without oscillation or mode collapse.

Across all experiments, the proposed oversmoothing metrics show a substantial reduction in excessive smoothing, which is also clearly reflected visually in the generated mel-spectrograms (Figure~\ref{fig:specs_comp}). 
The slight increase in L1/L2/SConv distances arises from the adversarial model producing more textured outputs that deviate from the reference at the fine-grained level---particularly in phase-dependent regions---while more closely matching the spectral richness of natural speech.

Given the small computational overhead of the adversarial discriminator, its architecture-agnostic nature, and the fact that it does not affect inference-time cost, we consider it a practical and effective addition to mel-spectrogram–based TTS systems and therefore employ it as the default configuration in the multi-speaker experiments.

\textbf{Cepstral Oversmoothing Trends.}
The evolution of HQER, CSlope, CCentroid, and CRoll95 confirms that adversarial training suppresses oversmoothing by restoring high-frequency cepstral contrast and improving the spectral slope and centroid structure of predicted mel-spectrograms. 
The early reduction of these metrics under adversarial training suggests faster convergence toward perceptually sharper spectra, whereas the slower decline observed in the baseline model reflects the gradual memorization of fine spectral detail rather than a genuine increase in spectral variability. 
Notably, the cepstral metrics continue to improve even after conventional reconstruction losses have plateaued, indicating that they provide a more sensitive measure of spectral fidelity during later stages of training. 
This behavior suggests that these metrics could serve as practical indicators for early stopping or for adaptively adjusting the weighting of adversarial objectives.

In summary, incorporating a lightweight adversarial spectrogram loss improves spectral realism and substantially reduces oversmoothing while preserving prosody and maintaining stable training behavior.

\subsection{Multi-Speaker Training}

The multi-speaker results demonstrate that adding synthetic XTTSv2 voices does not introduce 
negative transfer: the original ASC speaker maintains essentially the same performance as in the 
single-speaker setting. This indicates that the shared acoustic model can accommodate additional 
speaker variation without compromising its representation of the real target voice.

The differences in convergence behavior across speakers align with their intrinsic prosodic 
characteristics. Speakers with low pitch variability are easier to model and achieve lower 
$f_{0}$-RMSE, whereas wide pitch ranges increase modeling difficulty. The higher HQER and 
CCentroid values observed for female speakers follow directly from their higher fundamental 
frequency, which shifts harmonic energy toward higher quefrencies; these offsets therefore reflect 
physiological differences rather than modeling artifacts.

Adding synthetic speakers broadens the prosodic space seen during training and has a mild 
regularization effect, helping the model remain robust without increasing oversmoothing. This 
makes synthetic multi-speaker augmentation a practical strategy when only one real Arabic speaker 
is available. Future work could examine how model performance scales with larger numbers of 
synthetic speakers and whether diminishing returns or degradation eventually occur.

\section{Conclusions} \label{sec:conclusion}
The experiments presented in this work establish reproducible Arabic TTS baselines and introduce cepstral-domain metrics (HQER, CSlope, CCentroid, CRoll95) as effective indicators of oversmoothing. 
Consistent trends across models, training objectives, and speaker setups confirm that these metrics correlate with spectral richness. 
The lightweight adversarial discriminator substantially improves realism without compromising stability, and synthetic XTTSv2 speakers enhance prosodic diversity. 
Together, these findings provide a transparent foundation for future Arabic TTS studies focused on mitigating oversmoothing and improving perceptual fidelity.
\textbf{Future work} will aim to mitigate the biases introduced by the regression predictors and to evaluate the proposed metrics on more efficient TTS models suitable for mobile deployment.

\newpage

\bibliographystyle{IEEEtran}
\bibliography{refs}

@article{Shen2017NaturalTS,
  title={Natural TTS Synthesis by Conditioning Wavenet on MEL Spectrogram Predictions},
  author={Jonathan Shen and Ruoming Pang and Ron J. Weiss and Mike Schuster and Navdeep Jaitly and Zongheng Yang and Z. Chen and Yu Zhang and Yuxuan Wang and R. J. Skerry-Ryan and Rif A. Saurous and Yannis Agiomyrgiannakis and Yonghui Wu},
  journal={2018 IEEE International Conference on Acoustics, Speech and Signal Processing (ICASSP)},
  year={2017},
  pages={4779-4783},
  url={https://api.semanticscholar.org/CorpusID:206742911}
}

@article{Lancucki2020FastpitchPT,
  title={Fastpitch: Parallel Text-to-Speech with Pitch Prediction},
  author={Adrian La'ncucki},
  journal={ICASSP 2021 - 2021 IEEE International Conference on Acoustics, Speech and Signal Processing (ICASSP)},
  year={2020},
  pages={6588-6592},
  url={https://api.semanticscholar.org/CorpusID:219635877}
}

@misc{casanova2024xtts,
      title={XTTS: a Massively Multilingual Zero-Shot Text-to-Speech Model}, 
      author={Edresson Casanova and Kelly Davis and Eren Gölge and Görkem Göknar and Iulian Gulea and Logan Hart and Aya Aljafari and Joshua Meyer and Reuben Morais and Samuel Olayemi and Julian Weber},
      year={2024},
      eprint={2406.04904},
      archivePrefix={arXiv},
      primaryClass={eess.AS},
      url={https://arxiv.org/abs/2406.04904}, 
}

@misc{yu2019freeformimageinpaintinggated,
      title={Free-Form Image Inpainting with Gated Convolution}, 
      author={Jiahui Yu and Zhe Lin and Jimei Yang and Xiaohui Shen and Xin Lu and Thomas Huang},
      year={2019},
      eprint={1806.03589},
      archivePrefix={arXiv},
      primaryClass={cs.CV},
      url={https://arxiv.org/abs/1806.03589}, 
}

@misc{badlani2021ttsalignmentrule,
      title={One TTS Alignment To Rule Them All}, 
      author={Rohan Badlani and Adrian Łancucki and Kevin J. Shih and Rafael Valle and Wei Ping and Bryan Catanzaro},
      year={2021},
      eprint={2108.10447},
      archivePrefix={arXiv},
      primaryClass={cs.SD},
      url={https://arxiv.org/abs/2108.10447}, 
}

@inproceedings{Shih2021RADTTSPF,
  title={RAD-TTS: Parallel Flow-Based TTS with Robust Alignment Learning and Diverse Synthesis},
  author={Kevin J. Shih and Rafael Valle and Rohan Badlani and Adrian Łańcucki and Wei Ping and Bryan Catanzaro},
  year={2021},
  url={https://api.semanticscholar.org/CorpusID:235421175}
}

@misc{kim2021vits,
      title={Conditional Variational Autoencoder with Adversarial Learning for End-to-End Text-to-Speech}, 
      author={Jaehyeon Kim and Jungil Kong and Juhee Son},
      year={2021},
      eprint={2106.06103},
      archivePrefix={arXiv},
      primaryClass={cs.SD},
      url={https://arxiv.org/abs/2106.06103}, 
}

@misc{ren2019fastspeech,
      title={FastSpeech: Fast, Robust and Controllable Text to Speech}, 
      author={Yi Ren and Yangjun Ruan and Xu Tan and Tao Qin and Sheng Zhao and Zhou Zhao and Tie-Yan Liu},
      year={2019},
      eprint={1905.09263},
      archivePrefix={arXiv},
      primaryClass={cs.CL},
      url={https://arxiv.org/abs/1905.09263}, 
}

@phdthesis{nawar2016phdthesis,
author = {Halabi, Nawar},
year = {2016},
month = {07},
pages = {},
title = {Modern standard Arabic phonetics for speech synthesis}
}

@book{rabiner1978digital,
  title     = {Digital Processing of Speech Signals},
  author    = {Rabiner, Lawrence R. and Schafer, Ronald W.},
  year      = {1978},
  publisher = {Prentice Hall},
  address   = {Englewood Cliffs, NJ},
  isbn      = {978-0132136037}
}

@techreport{slaney1998auditory,
  author       = {Slaney, Malcolm},
  title        = {Auditory Toolbox},
  institution  = {Interval Research Corporation},
  number       = {Technical Report 1998-010},
  year         = {1998},
  address      = {Palo Alto, CA},
  url          = {https://engineering.purdue.edu/~malcolm/interval/1998-010/},
}

@article{Sakoe1978DynamicPA,
  title={Dynamic programming algorithm optimization for spoken word recognition},
  author={Hiroaki Sakoe},
  journal={IEEE Transactions on Acoustics, Speech, and Signal Processing},
  year={1978},
  volume={26},
  pages={159-165},
}

@Inbook{Müller2007,
  title="Dynamic Time Warping",
  bookTitle="Information Retrieval for Music and Motion",
  year="2007",
  publisher="Springer Berlin Heidelberg",
  address="Berlin, Heidelberg",
  pages="69--84",
  isbn="978-3-540-74048-3",
  doi="10.1007/978-3-540-74048-3_4",
  url="https://doi.org/10.1007/978-3-540-74048-3_4"
}

@book{hastie2009elements,
  title={The Elements of Statistical Learning},
  author={Hastie, Trevor and Tibshirani, Robert and Friedman, Jerome},
  year={2009},
  publisher={Springer}
}

@article{hillenbrand1994cpp,
author = {Hillenbrand, James and Cleveland, Ronald and Erickson, Robert},
year = {1994},
month = {08},
pages = {769-778},
title = {Acoustic Correlates of Breathy Vocal Quality},
volume = {37},
journal = {Journal of Speech, Language, and Hearing Research},
doi = {10.1044/jshr.3704.769}
}

@misc{Barqawi2017,
  title={Shakkala, Arabic text vocalization},
  author={Barqawi, Zerrouki},
  url={https://github.com/Barqawiz/Shakkala},
  year={2017}
}

@inproceedings{Fadel_2019,
   title={Neural Arabic Text Diacritization: State of the Art Results and a Novel Approach for Machine Translation},
   url={http://dx.doi.org/10.18653/v1/D19-5229},
   DOI={10.18653/v1/d19-5229},
   booktitle={Proceedings of the 6th Workshop on Asian Translation},
   publisher={Association for Computational Linguistics},
   author={Fadel, Ali and Tuffaha, Ibraheem and Al-Jawarneh, Bara’ and Al-Ayyoub, Mahmoud},
   year={2019},
   pages={215–225} 
}

@misc{alasmary2024cattcharacterbasedarabictashkeel,
      title={CATT: Character-based Arabic Tashkeel Transformer}, 
      author={Faris Alasmary and Orjuwan Zaafarani and Ahmad Ghannam},
      year={2024},
      eprint={2407.03236},
      archivePrefix={arXiv},
      primaryClass={cs.CL},
      url={https://arxiv.org/abs/2407.03236}, 
}

@misc{miyato2018spectralnormalization,
      title={Spectral Normalization for Generative Adversarial Networks}, 
      author={Takeru Miyato and Toshiki Kataoka and Masanori Koyama and Yuichi Yoshida},
      year={2018},
      eprint={1802.05957},
      archivePrefix={arXiv},
      primaryClass={cs.LG},
      url={https://arxiv.org/abs/1802.05957}, 
}

@misc{mao2017squaresgenerativeadversarialnetworks,
      title={Least Squares Generative Adversarial Networks}, 
      author={Xudong Mao and Qing Li and Haoran Xie and Raymond Y. K. Lau and Zhen Wang and Stephen Paul Smolley},
      year={2017},
      eprint={1611.04076},
      archivePrefix={arXiv},
      primaryClass={cs.CV},
      url={https://arxiv.org/abs/1611.04076}, 
}

@misc{loshchilov2019adamw,
      title={Decoupled Weight Decay Regularization}, 
      author={Ilya Loshchilov and Frank Hutter},
      year={2019},
      eprint={1711.05101},
      archivePrefix={arXiv},
      primaryClass={cs.LG},
      url={https://arxiv.org/abs/1711.05101}, 
}

@misc{kong2020hifigangenerativeadversarialnetworks,
      title={HiFi-GAN: Generative Adversarial Networks for Efficient and High Fidelity Speech Synthesis}, 
      author={Jungil Kong and Jaehyeon Kim and Jaekyoung Bae},
      year={2020},
      eprint={2010.05646},
      archivePrefix={arXiv},
      primaryClass={cs.SD},
      url={https://arxiv.org/abs/2010.05646}, 
}

@misc{siuzdak2024vocosclosinggaptimedomain,
      title={Vocos: Closing the gap between time-domain and Fourier-based neural vocoders for high-quality audio synthesis}, 
      author={Hubert Siuzdak},
      year={2024},
      eprint={2306.00814},
      archivePrefix={arXiv},
      primaryClass={cs.SD},
      url={https://arxiv.org/abs/2306.00814}, 
}

@misc{lin2018focallossdenseobject,
      title={Focal Loss for Dense Object Detection}, 
      author={Tsung-Yi Lin and Priya Goyal and Ross Girshick and Kaiming He and Piotr Dollár},
      year={2018},
      eprint={1708.02002},
      archivePrefix={arXiv},
      primaryClass={cs.CV},
      url={https://arxiv.org/abs/1708.02002}, 
}

@inproceedings{mauch2014pyin,
  author    = {Mauch, Matthias and Dixon, Simon},
  title     = {p{YIN}: A Fundamental Frequency Estimator Using Probabilistic Threshold Distributions},
  booktitle = {Proceedings of the IEEE International Conference on Acoustics, Speech and Signal Processing (ICASSP)},
  pages     = {659--663},
  year      = {2014},
  doi       = {10.1109/ICASSP.2014.6853678}
}

@manual{boersma2024praat,
  author       = {Boersma, Paul and Weenink, David},
  title        = {Praat: Doing Phonetics by Computer},
  organization = {Phonetic Sciences, University of Amsterdam},
  year         = {2024},
  note         = {Version 6.x},
  url          = {http://www.praat.org/}
}

@article{mann1947test,
  author    = {Mann, Henry B. and Whitney, Donald R.},
  title     = {On a Test of Whether One of Two Random Variables is Stochastically Larger than the Other},
  journal   = {The Annals of Mathematical Statistics},
  volume    = {18},
  number    = {1},
  pages     = {50--60},
  year      = {1947},
  doi       = {10.1214/aoms/1177730491}
}

@article{Arik_2019,
   title={Fast Spectrogram Inversion Using Multi-Head Convolutional Neural Networks},
   volume={26},
   ISSN={1558-2361},
   url={http://dx.doi.org/10.1109/LSP.2018.2880284},
   DOI={10.1109/lsp.2018.2880284},
   number={1},
   journal={IEEE Signal Processing Letters},
   publisher={Institute of Electrical and Electronics Engineers (IEEE)},
   author={Arik, Sercan O. and Jun, Heewoo and Diamos, Gregory},
   year={2019},
   month=jan, pages={94–98} }

\appendix

\newpage
\section{Curves}

\subsection{Multi-speaker}

\begin{figure}[h!]
\centering 
\begin{subfigure}{0.85\linewidth}
  \caption{Mel spectrogram distances}
  \includegraphics[width=\linewidth]{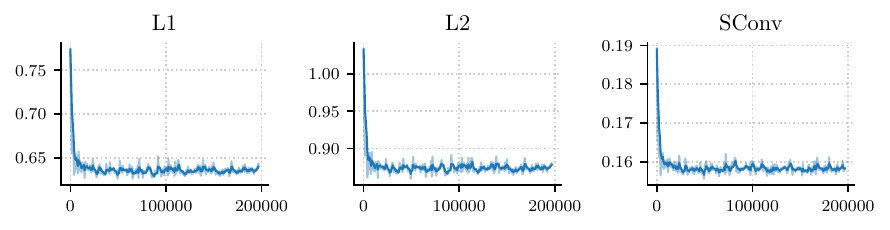}  
\end{subfigure}

\vspace{0.4em} 
\begin{subfigure}{0.85\linewidth}
  \caption{Pitch metrics \& Speaking rate}
  \includegraphics[width=\linewidth]{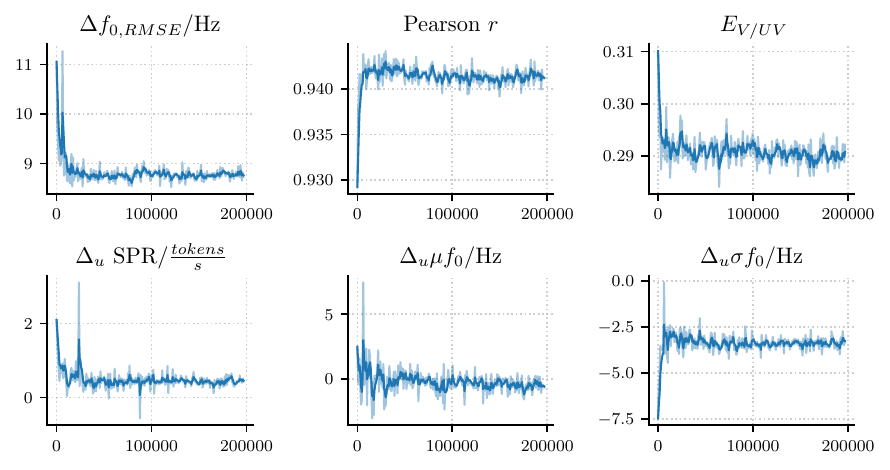}  
\end{subfigure}

\vspace{0.4em}
\begin{subfigure}{0.85\linewidth}
  \caption{Oversmoothing metrics}
  \includegraphics[width=\linewidth]{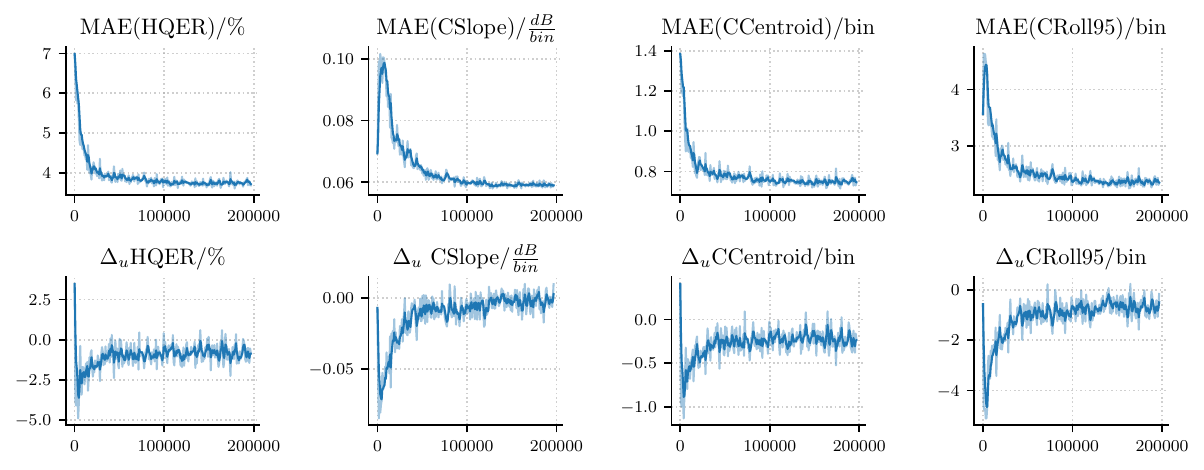}  
\end{subfigure}

\caption{Evaluation metric curves averaged across all speakers (S0-S3). 
Darker curves show an EMA-smoothed version with a smoothing factor of 0.6.}
\label{fig:curves-ms-avg}
\end{figure}

\newpage

\section{Metric Definitions}

\begin{table}[h!]
\centering
\caption{Oversmoothing metrics derived from the mel-cepstrogram. 
$\Delta$ denotes (pred -- ref).}
\renewcommand{\arraystretch}{1.2}
\begin{tabular}{p{2cm} p{7.5cm} p{5cm}}
\toprule
\textbf{Metric} & \textbf{Definition \& Interpretation} & \textbf{$\Delta$ (pred--ref)} \\
\midrule
HQER & Ratio of high- to total quefrency energy (excluding DC). High values indicate more fine spectral detail; low values indicate smoothing. & Negative $\Delta$ $\to$ more smoothing; positive $\Delta$ $\to$ more detail than reference. \\
\addlinespace
CSlope & Linear regression slope of $\log$-power vs.\ quefrency. Steeper negative slope means faster decay and less detail; flatter slope means more detail preserved. & More negative $\Delta$ $\to$ stronger smoothing; less negative $\Delta$ $\to$ closer to reference. \\
\addlinespace
CCentroid & Energy-weighted mean quefrency (normalized). Higher values mean energy at higher quefrencies (sharper); lower values mean energy collapsed at low quefrencies (smoother). & Negative $\Delta$ $\to$ smoother than reference; positive $\Delta$ $\to$ sharper than reference. \\
\addlinespace
CRoll95 & Quefrency index where 95\% of cumulative energy is reached. Larger values indicate broader spectrum and more detail; smaller values mean early cutoff and smoothing. & Negative $\Delta$ $\to$ more smoothing; positive $\Delta$ $\to$ more detail than reference. \\
\bottomrule
\end{tabular}
\end{table}

\paragraph{Pitch Metrics}

Let $f_0(m)$ denote the extracted frame-level pitch contour for a reference waveform 
$s_{\text{ref}}$ and $\hat{f}_0(m)$ the contour for the synthesized waveform 
$s_{\text{syn}}$, with $m=1,\dots,M$ denoting the frame index. 
Frames that are unvoiced are treated as undefined (NaN) and excluded 
from correlation and error metrics. 
\\\\
The \textbf{root mean squared error (RMSE)} of pitch is given by
\begin{align}
  \Delta f_{0,\text{RMSE}} 
    &= \sqrt{\frac{1}{N}\sum_{m \in \mathcal{V}} 
       \big(\hat{f}_0(m) - f_0(m)\big)^2},
\end{align}
where $\mathcal{V}$ is the set of frames voiced in both reference 
and synthesized signals, and $N = |\mathcal{V}|$ is their count. 
This measures the absolute framewise deviation in Hertz. 
\\\\
The \textbf{Pearson correlation coefficient} between the two contours is
\begin{align}
  r(f_0,\hat{f}_0) 
    &= \frac{\sum_{m \in \mathcal{V}} 
          \big(f_0(m)-\mu(f_0)\big)\,
          \big(\hat{f}_0(m)-\mu(\hat{f}_0)\big)}
          {\sqrt{\sum_{m \in \mathcal{V}} \big(f_0(m)-\mu(f_0)\big)^2}\,
           \sqrt{\sum_{m \in \mathcal{V}} \big(\hat{f}_0(m)-\mu(\hat{f}_0)\big)^2}},
\end{align}
where $\mu(\cdot)$ denotes the mean across the voiced set $\mathcal{V}$. 
This quantifies how well the pitch trajectories co-vary, independent of scale. 
\\\\
The \textbf{voiced/unvoiced (V/UV) error rate} is defined as
\begin{align}
  E_{\text{V/UV}} 
    &= \frac{1}{M} \sum_{m=1}^{M} 
       \mathbf{1}\!\left[\text{voiced}\big(f_0(m)\big) 
       \neq \text{voiced}\big(\hat{f}_0(m)\big)\right],
\end{align}
where $\mathbf{1}[\cdot]$ is the indicator function and 
$\text{voiced}(\cdot)$ returns a Boolean voiced/unvoiced decision.

\end{document}